\documentclass[review,number,sort&compress]{elsarticle}
\usepackage{amssymb}
\usepackage{upgreek}
\usepackage{lineno}

\begin{document}

\begin{frontmatter}



 \title{Charge collection properties of irradiated depleted CMOS pixel test structures\tnoteref{t1}}
 \tnotetext[t1]{The work was partly done in the framework of the RD50 collaboration.}

 \author[ijs]{I.~Mandi\' c\corref{cor1}}
 \ead{igor.mandic@ijs.si}

 \author[ijs]{V. ~Cindro}
 \author[ijs]{A. ~Gori\v sek}
 \author[ijs]{B. ~Hiti}
 \author[ijs]{G. ~Kramberger}
\author[ijs]{M. ~Zavrtanik}
\author[ijs,fmf]{M. ~Miku\v z}
\author[bonn]{T. ~Hemperek}

\cortext[cor1]{Corresponding author}

\address[ijs]{Jo\v zef Stefan Institute, Jamova 39, Ljubljana, Slovenia}
\address[fmf]{University of Ljubljana, Faculty of Mathematics and Physics, Jadranska 19, Ljubljana, Slovenia}
\address[bonn]{Physikalisches Institut, Universit\" at Bonn, Nu\ss allee 12, 53115 Bonn, Germany}


\begin{abstract}

Edge-TCT and charge collection measurements with passive test structures made in LFoundry 150 nm CMOS process on p-type
substrate with initial resistivity of over 3 k$\Omega$cm are presented. Measurements were made before and
after irradiation with reactor neutrons up to 2$\cdot$10$^{15}$ n$_{\mathrm{eq}}$/cm$^2$.
Two sets of devices were investigated: unthinned (700 $\upmu$m) with substrate biased through the implant on top and
thinned (200 $\upmu$m) with processed and metallised backplane.

Depletion depth was estimated with Edge-TCT and collected charge was measured with $^{90}$Sr source using an external
amplifier with 25 ns shaping time. Depletion depth at given bias voltage decreased with increasing neutron fluence but
it was still larger than 70 $\upmu$m at 250 V after the highest fluence. After irradiation much higher collected charge was
measured with thinned detectors with processed backplane although the same depletion depth was observed with Edge-TCT.
Most probable value of collected charge of over 5000 electrons was measured also after irradiation to 2$\cdot$10$^{15}$ n$_{\mathrm{eq}}$/cm$^2$. This is sufficient
to ensure successful operation of these detectors at the outer layer of the pixel detector in the ATLAS experiment at the upgraded HL-LHC.
\end{abstract}


\end{frontmatter}

%%\linenumbers

\section{Introduction}
\label{intro}

The upgrade of the Large Hadron Collider (LHC) to High Luminosity LHC (HL-LHC) \cite{Fabiola} foreseen in the next decade will significantly increase
the rate of proton collisions at the interaction point. As a result the number of charged tracks generated in each bunch crossing will increase leading
to a harsher radiation environment  \cite{radlevel}. This will require replacement of the present pixel detectors in general purpose experiments like ATLAS \cite{ATLAS,ATLASpix} and CMS \cite{CMSpix} because their granularity and radiation hardness is not sufficient for HL-LHC.
In the upgraded ATLAS experiment \cite{LOT} large areas will be covered with silicon detectors so cost,
production time and complexity of assembly require attention in addition to tracking performance to ensure the success of the project. All these could be greatly improved with monolithic depleted pixel detectors produced in a commercial CMOS process. Charge collection from the depletion region is necessary for sufficient speed and radiation hardness for application at high luminosity hadron colliders. Monolithic detector approach
would greatly simplify the assembly, and production in an industrial CMOS process on large wafers in high volume foundries would speed up the production and
lower the cost.

Research of this detector technology for application at LHC was initiated by developments in HV-CMOS process \cite{peric} and has become very intensive recently. Promising results showing sufficient radiation hardness of depleted CMOS detectors from various designs and producers were published in a large number of publications - references
 \cite{Pohl}- \cite{FirstLF} 
%%\cite{Pohl, Pernegger, cavallaro, Fernandez, JinstChess, Fadeyev, kanisauskas, hiti, hemperek, FirstLF}
represent just a few more recent ones. CMOS technology is being investigated as an option for the outermost layer of the pixel detector in the upgraded ATLAS experiment \cite{LOT} where expected displacement damage in silicon caused by energetic hadrons will be equivalent to the damage caused by a fluence of 2$\cdot$10$^{15}$ 1 MeV neutrons per cm$^2$ \cite{radlevel}.

One of the investigated versions of depleted CMOS detectors is produced by LFoundry
\cite{Lfoundry} - \cite{hirono}
%\cite{Lfoundry, kinishita, Rymaszevski, hirono}
in a 150 nm process on a p-type substrates with initial resistivities exceeding 2 k$\Omega$cm and typically in the range between 4 and 5 k$\Omega$cm \cite{Lfoundry}. Such a high resistivity material ensures a large depletion depth already at moderate bias voltages.
An irradiation study using Edge-TCT technique with passive test structures manufactured by LFoundry was published in \cite{FirstLF} and it
was shown that depletion depth of over 50 $\upmu$m is achieved at 120 V bias also after irradiation with 8$\cdot$10$^{15}$ n$_{\mathrm{eq}}$/cm$^2$.
In this paper we report on measurements with a new set of samples produced in the same process. Samples were characterised using
Edge-TCT technique similar as described in \cite{FirstLF}. In addition, collected charge deposited in the detector by a passage of a fast
electron from $^{90}$Sr source was measured and compared to the value estimated from the depletion depth measured by Edge-TCT.
Two sets of devices were investigated: devices from an unthinned wafer with
substrate biased through the implant on top and devices from a thinned wafer with processed and metallised backplane for substrate contact.

\section{Samples and irradiation}
\label{samples}

\begin{figure}[!hbt]
\centering
a)   \includegraphics[width=0.7\textwidth]{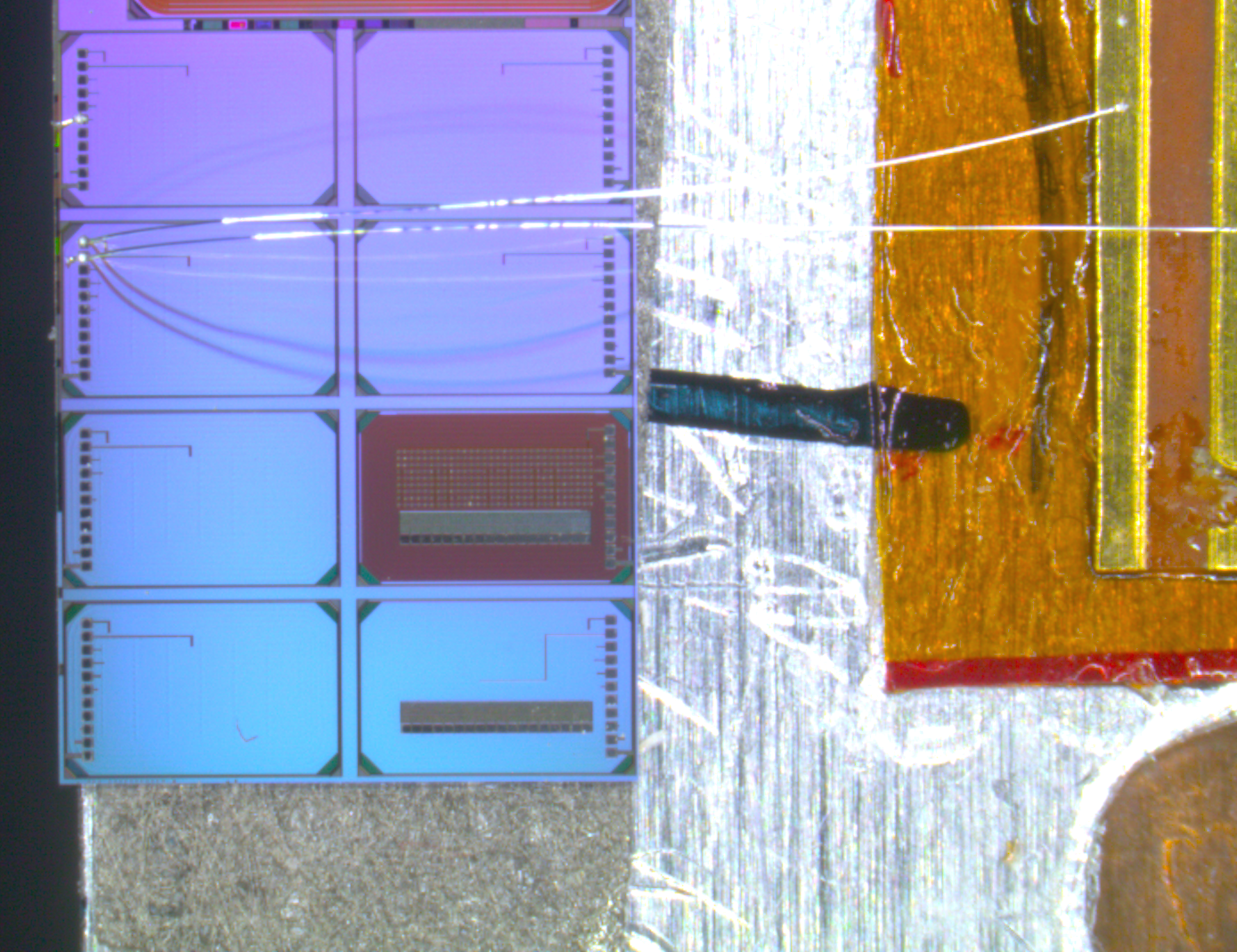}
   \begin{tabular}{cc} 
     \includegraphics[width=0.6\textwidth]{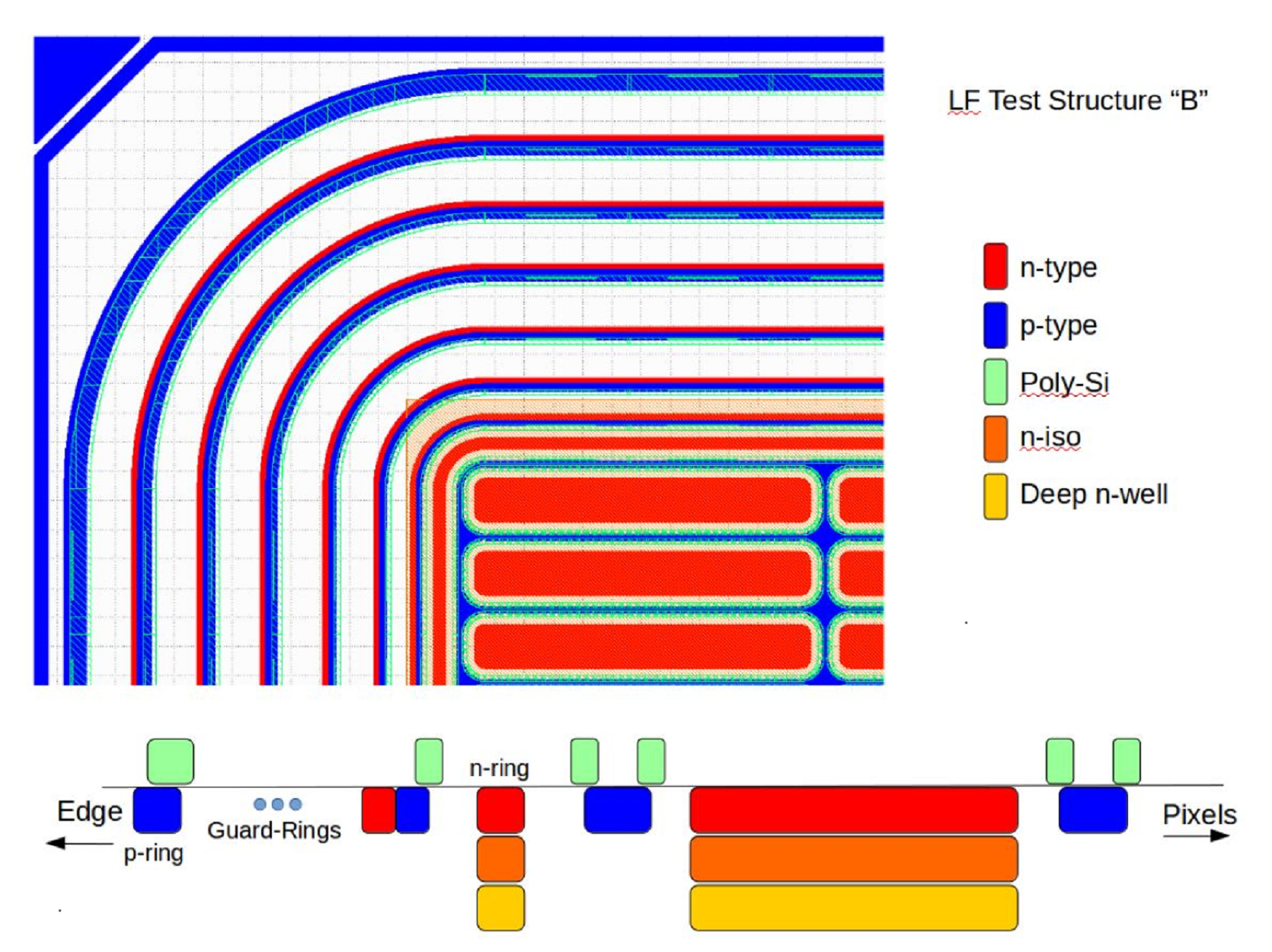} & \includegraphics[width=0.35\textwidth]{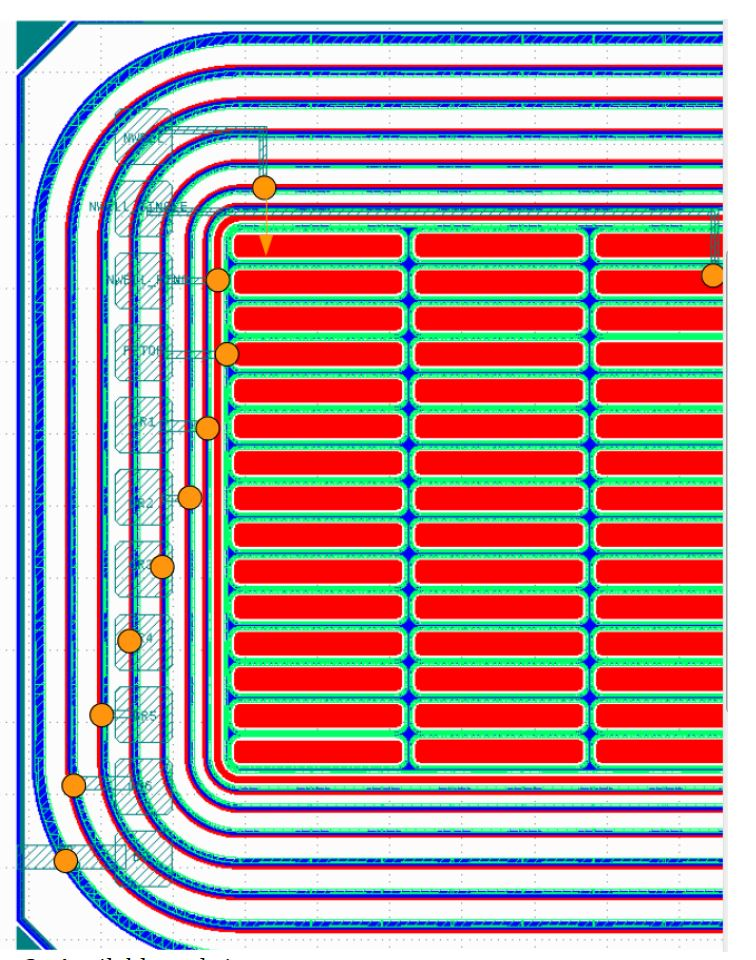} \\
     b) & c)
   \end{tabular}
   \caption{a) Photo of the thinned LFoundry chip biased through the backplane prepared for measurement. b) Schematic cross section of test structure B. c) Layout of (part of) structure B with bonding pads.}
  \label{LFstruct_photo}
\end{figure}

The photo in Fig. \ref{LFstruct_photo}a) shows the LFoundry chip with several test structures. Measurements were made with test structure B
schematically shown in drawings in Fig. \ref{LFstruct_photo}b and \ref{LFstruct_photo}c.
Structure B is an array of 15 $\times$ 6 passive n$^+$p pixels with a pixel size of 50$\times$520 $\upmu$m$^2$.
The structure has 11 bonding pads: a pad for all pixels except one connected together and a separate bonding pad
for this single pixel (see  Fig. \ref{LFstruct_photo}c)). The remaining bonding pads connect to the p-type implants between n-wells in the pixel area,
the n-ring surrounding the pixels and several guard
rings. The outermost p-type implant ring also has a separate bonding pad and was used to bias the substrate.

Chips are produced on 700 $\upmu$m thick wafers and unthinned samples with no backplane processing were diced from the wafer. A set of
samples was taken from a wafer thinned to ~200 $\upmu$m with processed and metallised backplane.

Before measurements each chip was fixed to an aluminium support using conductive glue
(see Fig. \ref{LFstruct_photo}). For unthinned samples the substrate was contacted via a wire bond to the p-type ring while for thinned samples contact to the substrate was achieved via the backplane
through the conductive glue.

Chips were irradiated with neutrons in the TRIGA reactor in Ljubljana \cite{Reactor1,Reactor2} to 1 MeV neutron equivalent
fluences ranging from 1$\cdot$10$^{13}$ n$_{\mathrm{eq}}$/cm$^2$ to  2$\cdot$10$^{15}$ n$_{\mathrm{eq}}$/cm$^2$. In the reactor, irradiation is performed by inserting the samples into the core through irradiation tubes. The maximal power of the TRIGA reactor in Ljubljana is 250 kW. At this power
1 MeV neutron equivalent flux in the irradiation channel is 1.5$\cdot$10$^{12}$ n$_{\mathrm{eq}}$/cm$^2$/s therefore the
fluence of 1$\cdot$10$^{14}$ n$_{\mathrm{eq}}$/cm$^2$ is reached in 65 s. Irradiation to lower fluences is performed at lower reactor power
to increase irradiation time and thus reduce irradiation time uncertainties due to insertion and extraction. Neutron flux in the irradiation channel
is periodically controlled by measurement of leakage current increase in dedicated silicon diodes irradiated in this irradiation
channel \cite{dosimetry}. Before measurements samples were annealed for 80 minutes at 60$^\circ$C.

\section{Experimental techniques}
\label{techniques}

Edge-TCT is a variant of Transient Current measurement Technique in which sub-nanosecond pulses of infra-red
($\lambda$ = 1064 nm) laser light are directed to the edge (i.e. laser beam runs parallel with the surface) of the investigated
device. Narrow laser beam with a diameter of less than 10 $\upmu$m is used so that charge carriers are released at a known depth
in the investigated device positioned in the beam with high precision moving stages. Edge-TCT method was first described in
\cite{Edge-TCT} and has by now become a standard tool for investigating radiation effects in silicon detectors. Measurements
within this work were carried out with an Edge-TCT system produced by Particulars \cite{Particulars}. Edge-TCT data taking and
analysis techniques are very similar to those described in \cite{FirstLF} and details can be found there. 

\begin{figure}[!hbt]
  \centering
   \includegraphics[width=0.6\textwidth]{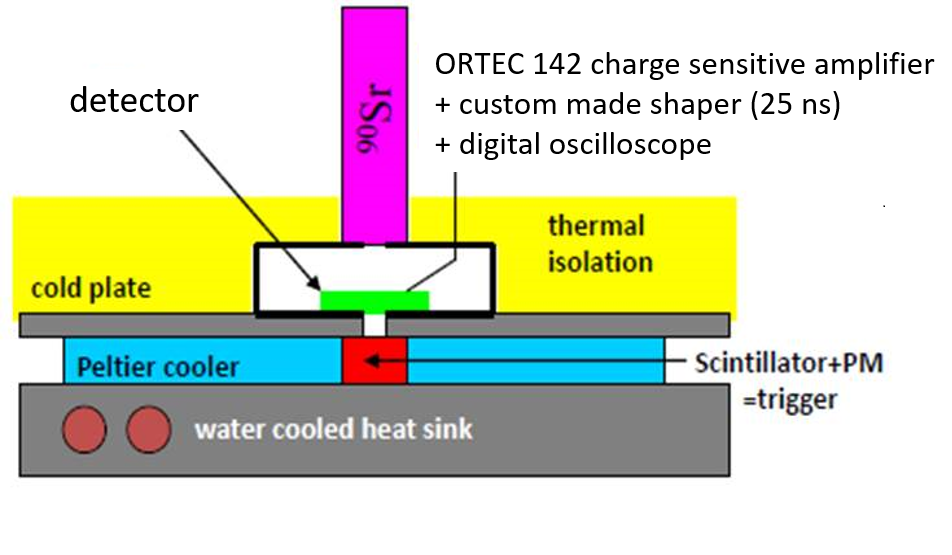}
  \caption{Scheme of experimental setup for charge collection measurements with $^{90}Sr$ source.}
  \label{setup}
\end{figure}

Charge collection measurements were made with the experimental setup shown in Fig. \ref{setup}. The DUT was placed between two aluminium collimators with 1 mm diameter holes with $^{90}$Sr source on one side and a small scintillator coupled to a photomultiplier on the other. Only electrons from the high energy end of the $^{90}$Sr spectrum have sufficient energy to pass the silicon chip and trigger the readout by depositing sufficient energy in the scintillator. Collimators minimize the contribution of electrons scattered by a large angle in the setup. 

This enables measurement of charge released in the detector by passage of a particle which is a close approximation of a 
Minimum Ionising Particle (MIP). But it should be mentioned that because of lower energy and larger scattering of electrons from $^{90}$Sr compared to MIPs measured collected charge will be of the order of 10\% larger than in the case of MIPs. 

When triggered by the photomultiplier digital oscilloscope records the waveform from the custom made shaping circuit with 25 ns
peaking time processing the signal from the Ortec 142 charge sensitive preamplifier. Active area of the structure B used in this work (see Fig. \ref{LFstruct_photo}b) is 0.75 mm x 3.1 mm which is a bit too small compared to collimator hole diameter to ensure a
sample with large fraction of waveforms with charge deposited in the structure B. So there is about 50\% of waveforms in the
sample where electrons from the source triggered the readout but did not deposit charge in the structure B.
Therefore collected charge could be estimated if peaks of measured distribution with and without charge signal in the structure could be clearly separated. This was the case if Most Probable Value (MPV) extracted from the fit of convolution of Landau and Gaussian distributions to the signal spectrum was larger than about 4000 electrons.

The system was calibrated by measuring the Most Probable Value (MPV) of collected charge of a standard
300 $\upmu$m thick fully depleted silicon detector and confirmed with 59.5 keV photons from $^{241}$Am source.
More detail about this measurement setup is in  \cite{JinstChess,SrSetup}.

\section{Measurements}
\label{measurements}

\subsection{Edge-TCT}

In Edge-TCT current pulses induced on readout electrodes by movement of charge carriers released in the detector by a short laser pulse are recorded and analysed. The time integral - integration time was 25 ns in this work - of the current pulse is defined as the charge. In Edge-TCT measurements shown in this work
single pixel (see Fig. \ref{LFstruct_photo}) was connected to the wide bandwidth amplifier while other pixels were connected to the same potential but not to the amplifier. High voltage was connected to the pixels and decoupled from readout by a Bias-T while substrate contact was
at ground potential.

\begin{figure}[!hbt]
  \centering
  a)\\
  \includegraphics[width=0.7\textwidth]{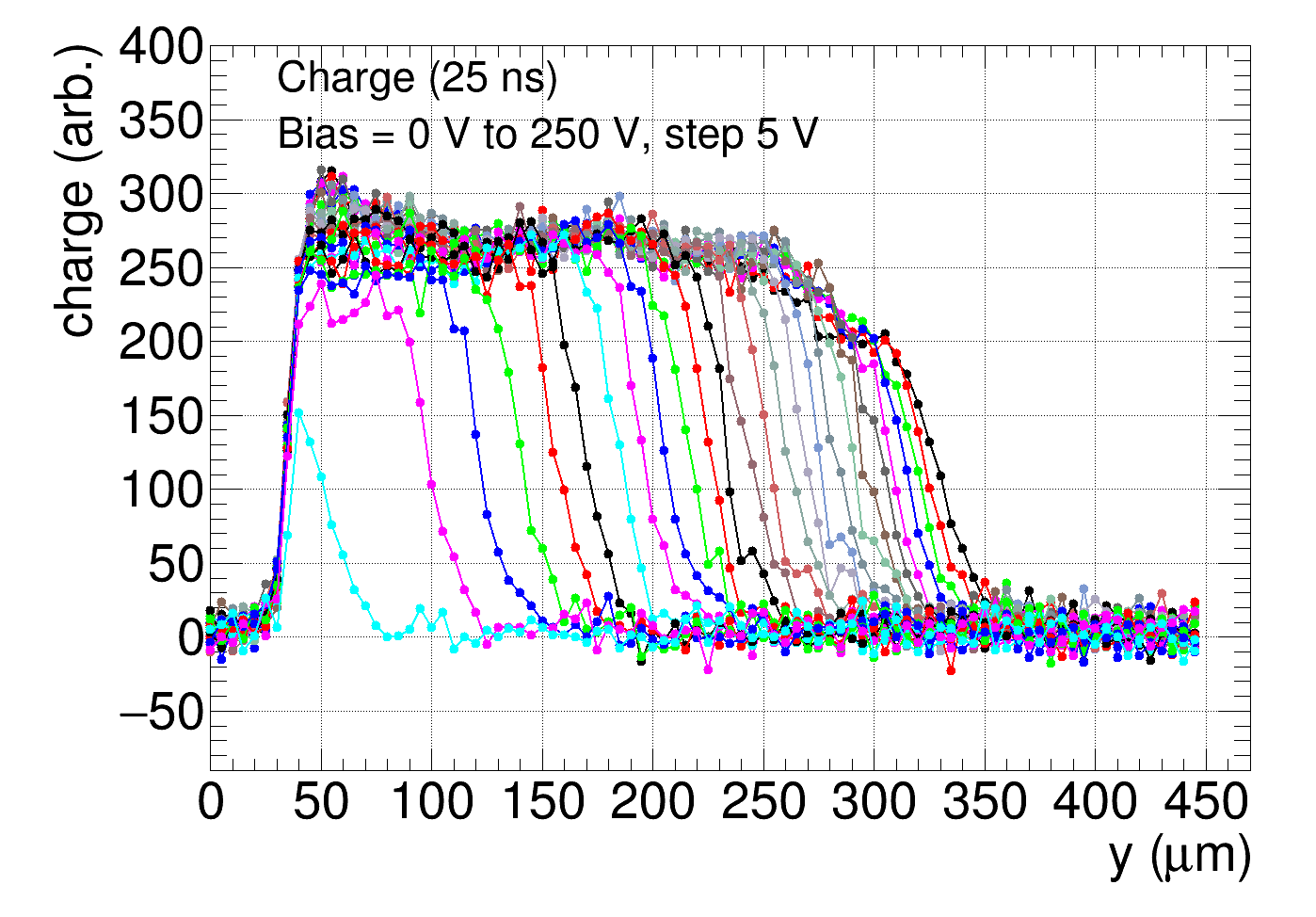}
  
  b) \\
  \includegraphics[width=0.7\textwidth]{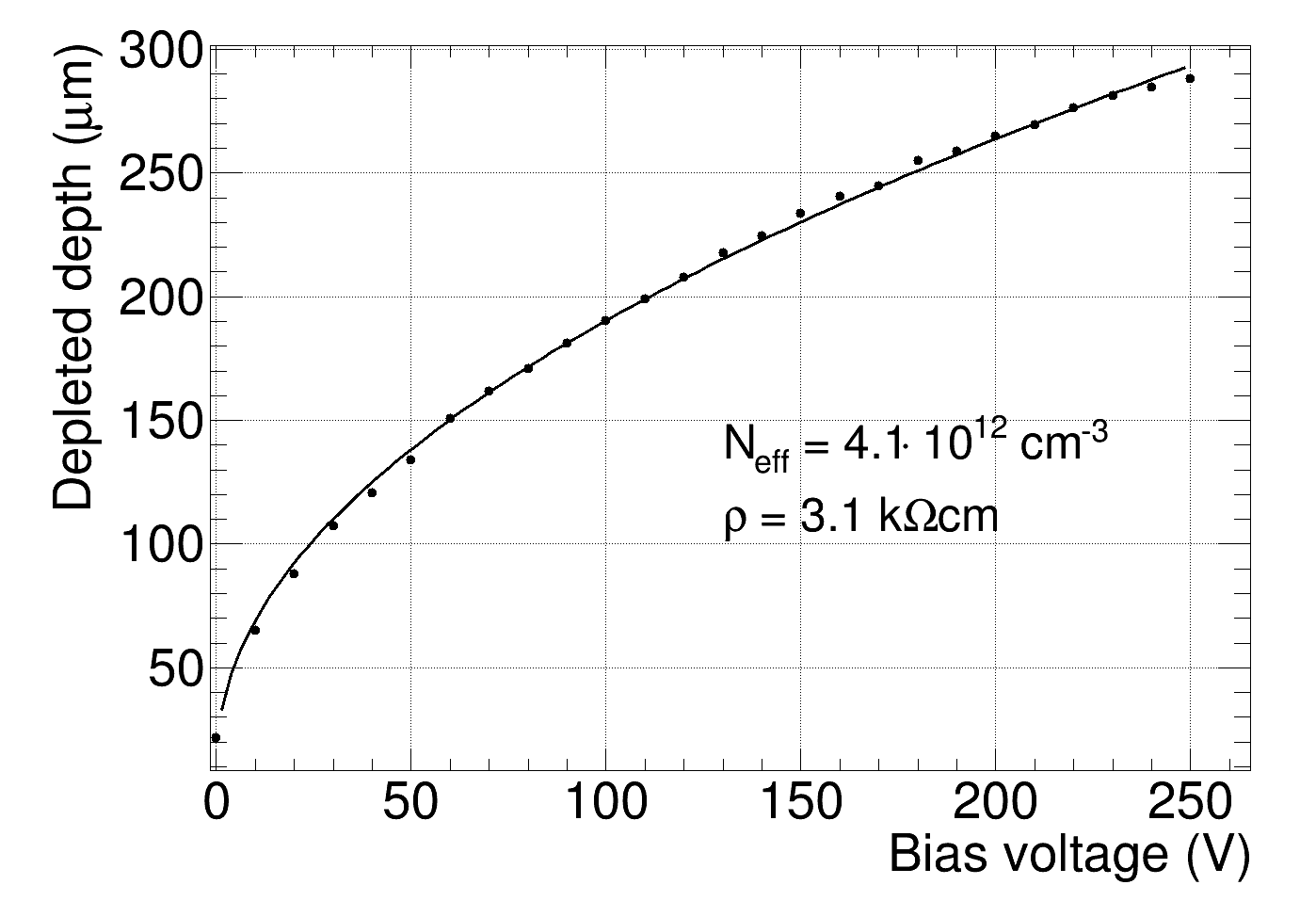} 
  \caption{Figure a): charge collection profiles for an unthinned device before irradiation at bias voltages from 0 to 250 V in 5 V steps. The narrowest profile was measured at 0 V. Figure b): FWHM of charge collection profiles vs. bias voltage.}
  \label{ETCTunirrad}
\end{figure}

\begin{figure}[!hbt]
  \centering
  \includegraphics[width=0.7\textwidth]{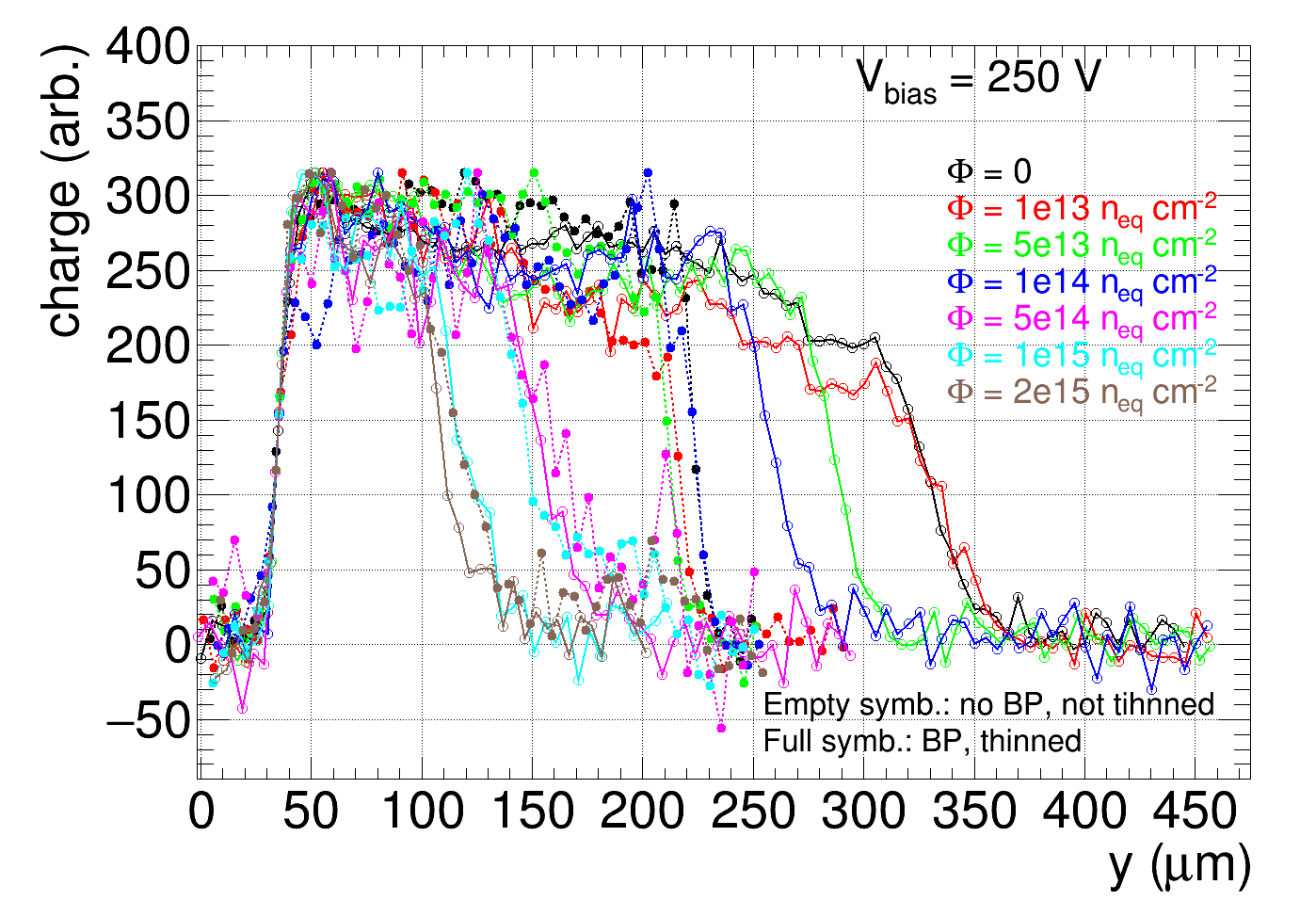} \\
a)\\
\includegraphics[width=0.7\textwidth]{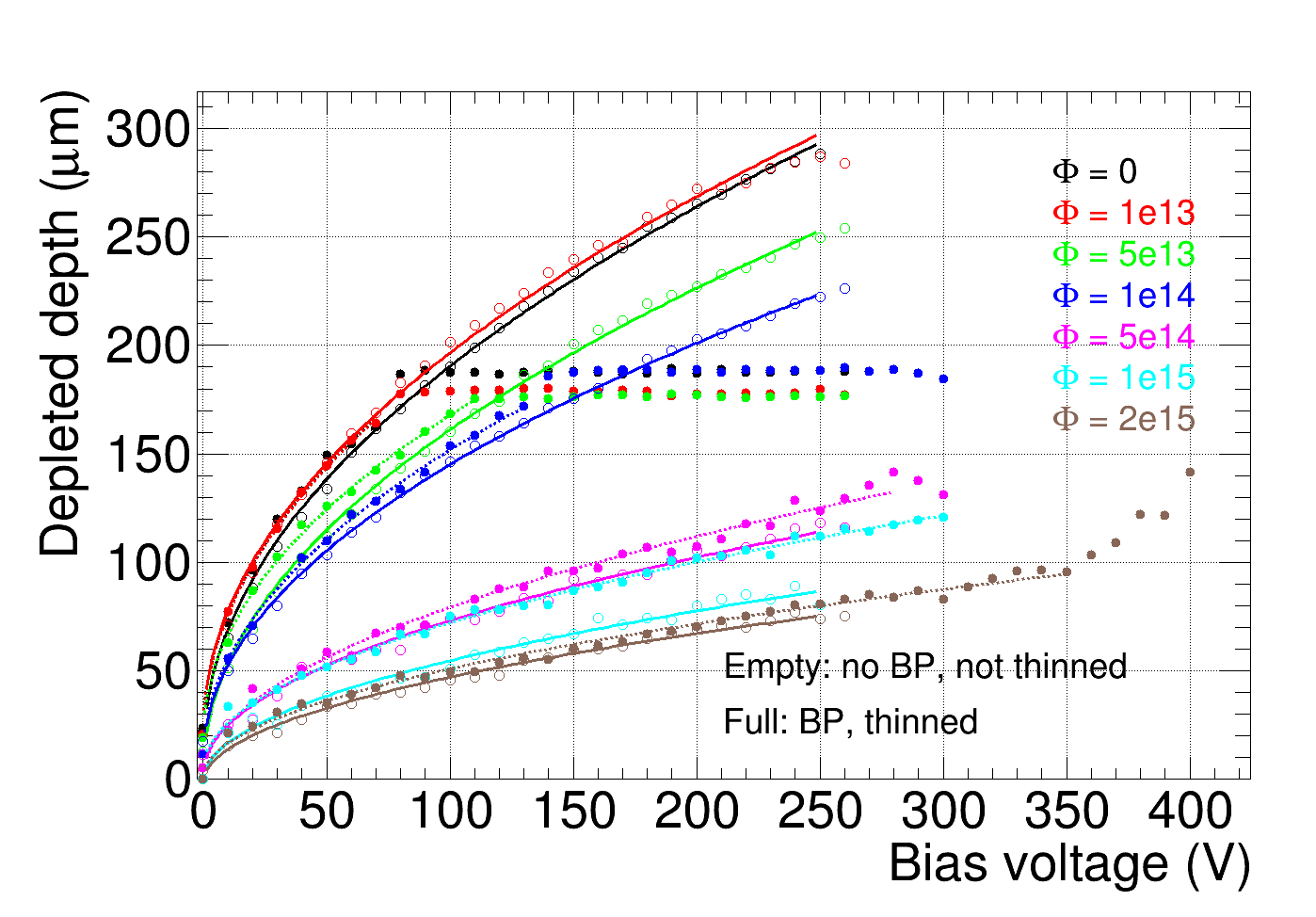} \\
b)
  \caption{Figure a): charge collection profiles at V$_{\mathrm{bias}}$ = 250 V and b) width of profiles vs. bias voltage at different fluences for thinned and unthinned devices. For each fluence a different device was measured.}
  \label{ETCTirrad}
\end{figure}

Figure \ref{ETCTunirrad}a shows charge as a function of substrate depth at an increasing bias voltage for an unthinned device before irradiation.  Charge is given in arbitrary units because the signal depends on laser power, laser beam focusing, edge surface quality etc... and thus cannot be compared between devices. It can be seen how the width of the charge collection profile, which is a measure of depletion depth, increases with bias voltage. In Fig. \ref{ETCTunirrad}b the FWHM of the profile is plotted vs. bias voltage and fitted with: 
\begin{equation}
  w(V_{\mathrm{bias}}) = w_0 + \sqrt{\frac{2\epsilon_r\epsilon_0}{e_0N_{\mathrm{eff}}}V_{\mathrm{bias}}}
  \label{fitfunc}
\end{equation}
where $N_{\mathrm{eff}}$ is the effective space charge concentration, $V_{\mathrm{bias}}$ the bias voltage, $e_0$ the elementary
charge, $\epsilon_0$ the dielectric constant and $\epsilon_r$ the relative permittivity of silicon. In the approximation of
abrupt junction
and uniform doping the parameter $w_0$ would be zero at zero bias voltage (neglecting the builtin voltage). However, it is
known \cite{Fernandez, JinstChess,FirstLF, Fernandez_1} that in Edge-TCT measurements at shallow depletion depths a significant offset is observed
due to a finite laser beam width, charge collected by diffusion and by laser beam reflections from the metallised surfaces.

We observe that the Eq. \ref{fitfunc} function fits the data in Fig. \ref{ETCTunirrad} well if  $N_{\mathrm{eff}}$ and  $w_0$ are free parameters. 
$N_{\mathrm{eff}}$ returned by the fit corresponds to an initial resistivity of 3 k$\Omega$cm, in agreement with spectification.

Figure \ref{ETCTirrad}a) shows charge collection profiles of thinned and unthinned devices after different neutron fluences at 250 V bias. For each irradiation fluence a different device was measured and the charge profiles
for different devices were rescaled to have equal maximal charge. It can be seen that for unthinned devices the charge profile gets narrower with increasing fluence. For thinned devices the profiles up to the fluence of 1$\cdot$10$^{14}$ n$_{\mathrm{eq}}$/cm$^2$ have similar widths of about 180 $\upmu$m because they are fully depleted at this bias voltage. It is also important to note that at higher fluences the profile widths for thinned and unthinned devices are similar (except at the fluence of 1$\cdot$10$^{15}$ n$_{\mathrm{eq}}$/cm$^2$ - the reason for this deviation is not understood).

In Fig \ref{ETCTirrad}b) charge profile widths are plotted versus bias voltage. It can nicely be seen that in thinned devices the depletion depth grows with bias voltage in accordance with equation Eq. \ref{fitfunc} until full depletion is reached. Before full depletion the dependence on bias voltage is very similar for thinned and unthinned devices (except for the already mentioned case of $\Phi_{\mathrm{eq}}$ = 1$\cdot$10$^{15}$ n$_{\mathrm{eq}}$/cm$^2$).
It can be seen that width at full depletion depths is 190 $\upmu$m for two fluences and 170 $\upmu$m for the other two. For each fluence different device was measured so this may be the consequence of variation of the thickness of active depth after thinning and backplane processing.

\begin{figure}[!hbt]
  \centering
  \includegraphics[width=0.7\textwidth]{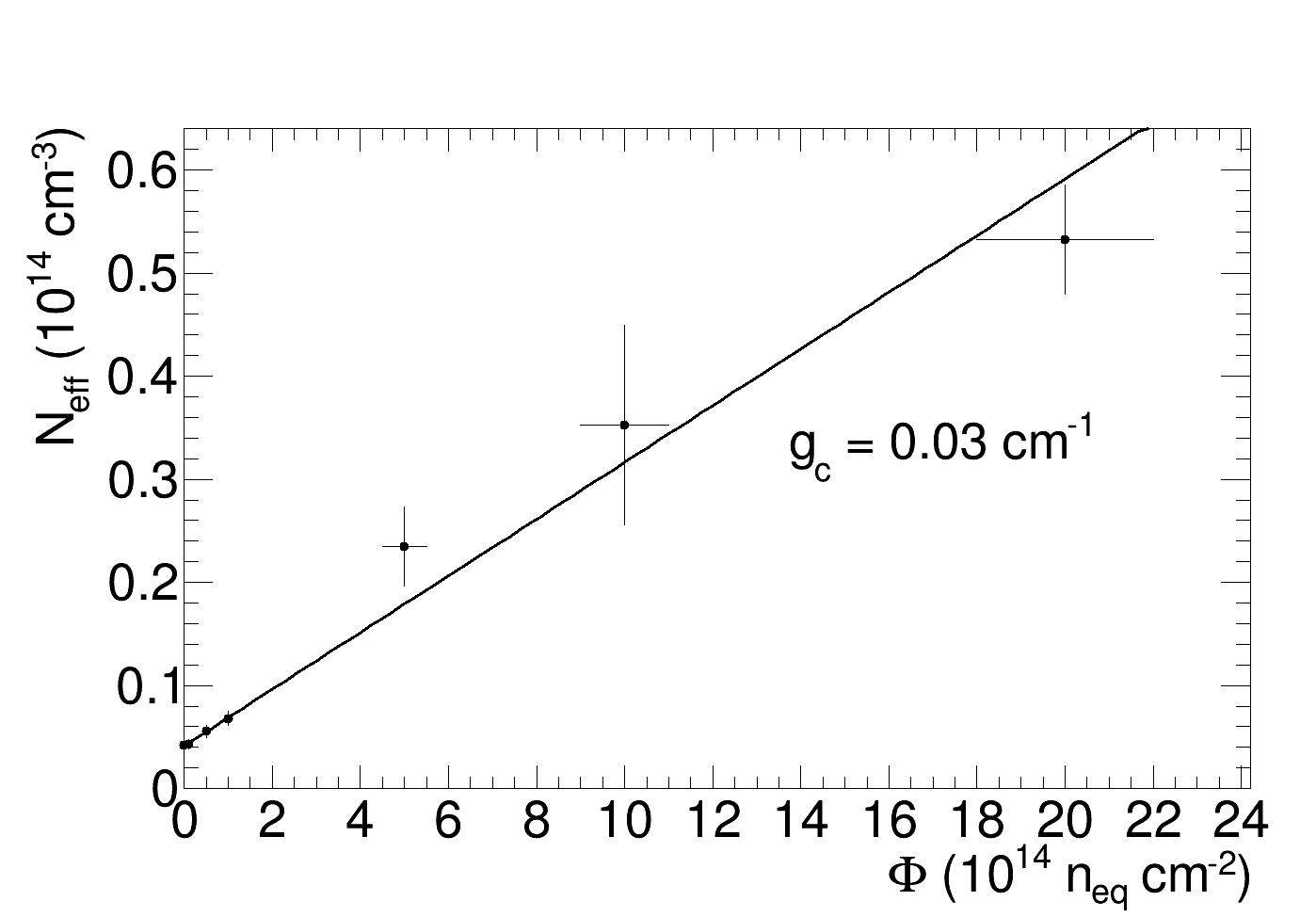}
  \caption{Effective doping concentration as a function of 1 MeV equivalent neutron fluence.}
  \label{NeffVsFlu}
\end{figure}
$N_{\mathrm{eff}}$ can be estimated from the fit to Eq. \ref{fitfunc} also after irradiation. Figure \ref{NeffVsFlu} shows $N_{\mathrm{eff}}$, the average of values from thinned and unthinned devices, as a function of neutron fluence and it can be seen that it increases linearly with fluence with slope $g_c \sim 0.03$ cm$^{-1}$. This is somewhat larger
than the stable damage introduction rate usually found in p-type silicon ($g_c \sim 0.02$ cm$^{-1}$ \cite{cindrop}) but it is consistent with a similar measurement with LFoundry samples from \cite{FirstLF}. No decrease of $N_{\mathrm{eff}}$ with fluence - a strong indication of initial acceptor removal - was observed in this work. In measurements in \cite{FirstLF} a decrease of $N_{\mathrm{eff}}$  was measured at 1$\cdot$10$^{13}$ n$_{\mathrm{eq}}$/cm$^2$  and 5$\cdot$10$^{13}$ n$_{\mathrm{eq}}$/cm$^2$ and acceptor removal parameters could be estimated. But it should be noted that the initial resistivity of samples measured in \cite{FirstLF} was
$\sim$2 k$\Omega$cm while for samples measured in this work it is over 3 k$\Omega$cm (see Fig. \ref{ETCTunirrad}). It may be expected, based on observations in \cite{JinstChess,LgadKrambi}, that the acceptor removal effects would be observable at lower fluences than measured in this work because of higher initial resistivity.

\subsection{Charge collection measurements}

\begin{figure}[!hbt]

  \begin{tabular}{cc}
    
  \includegraphics[width=0.5\textwidth]{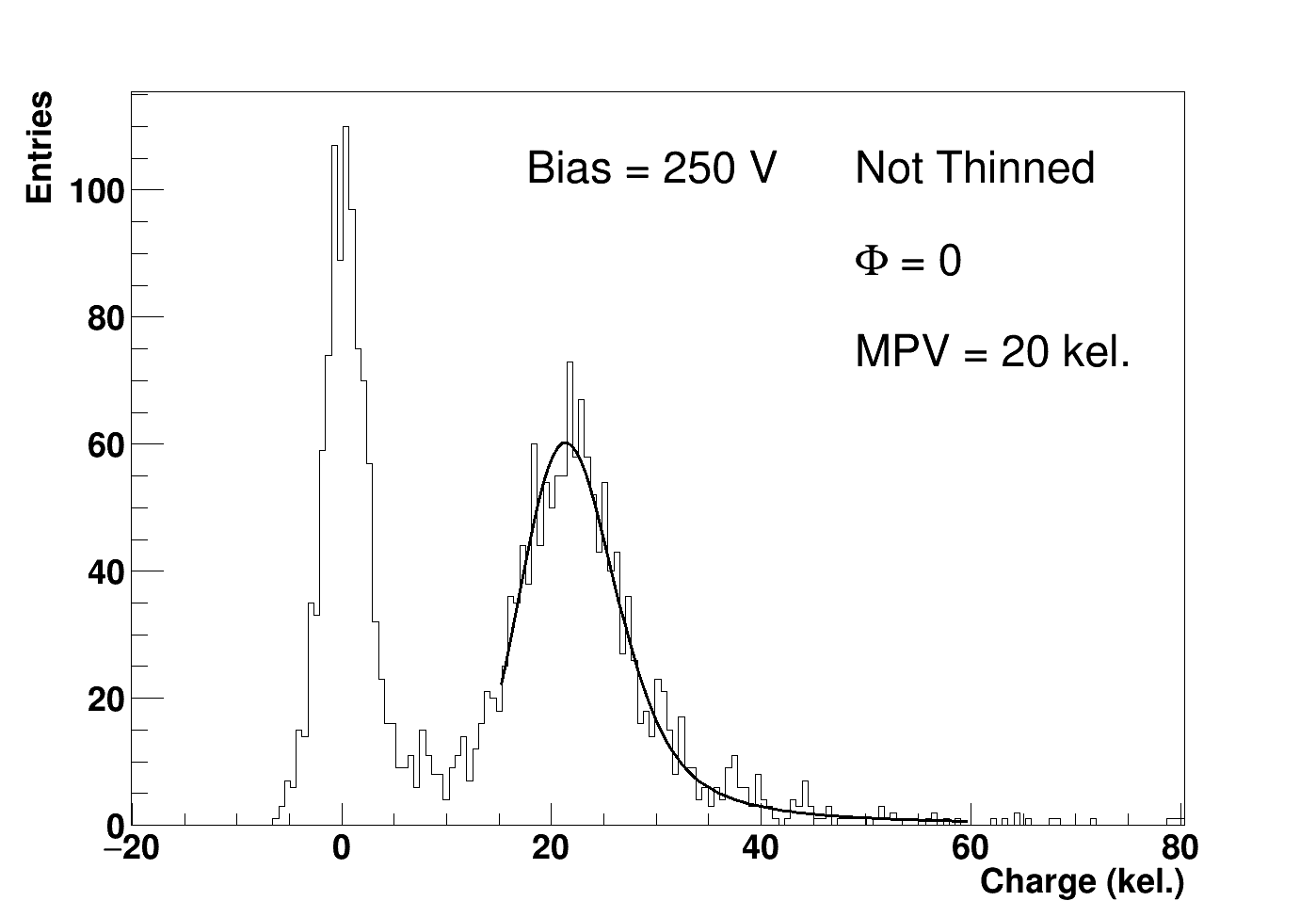} & \includegraphics[width=0.5\textwidth]{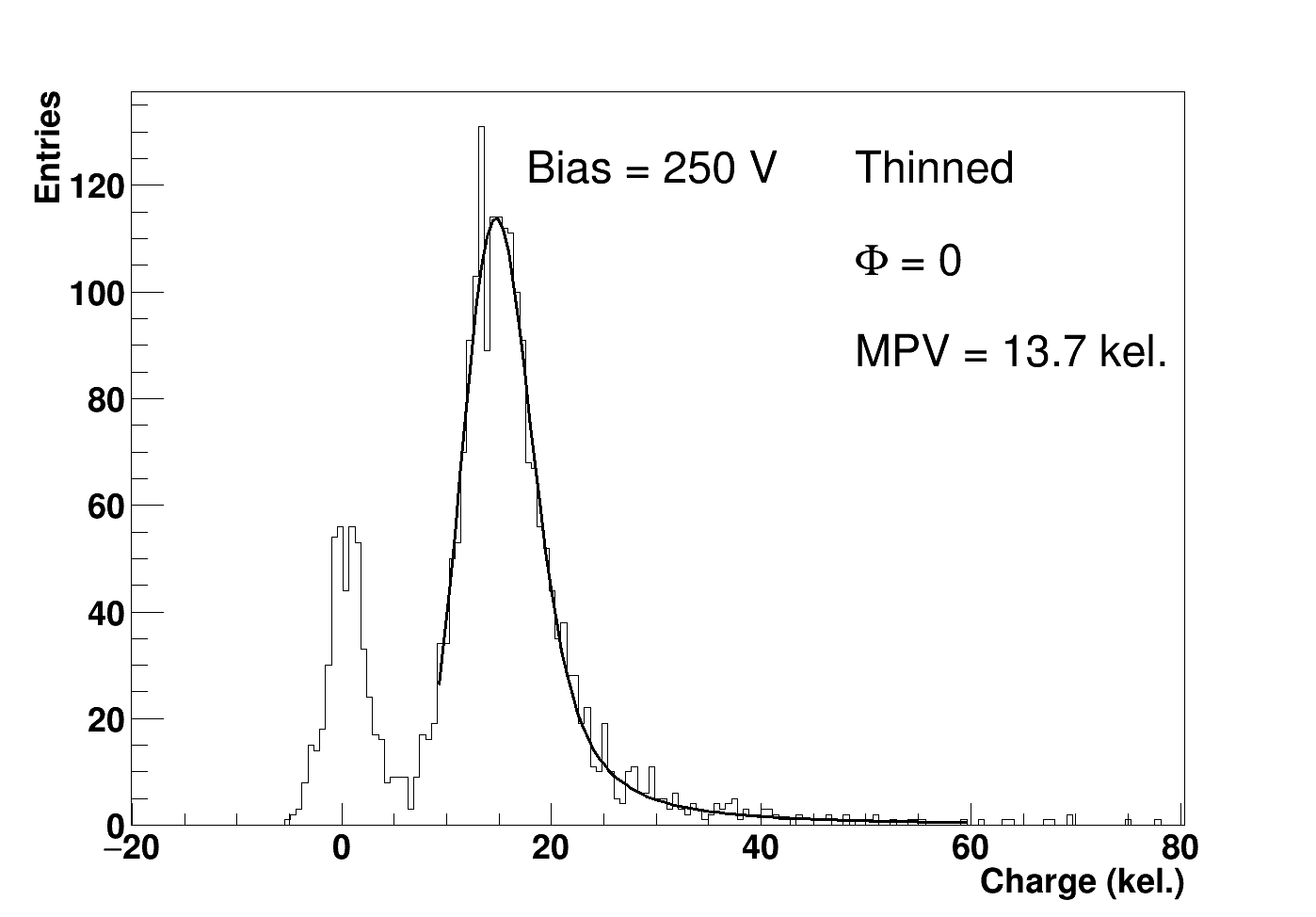} \\

  a) & b) \\

  \includegraphics[width=0.5\textwidth]{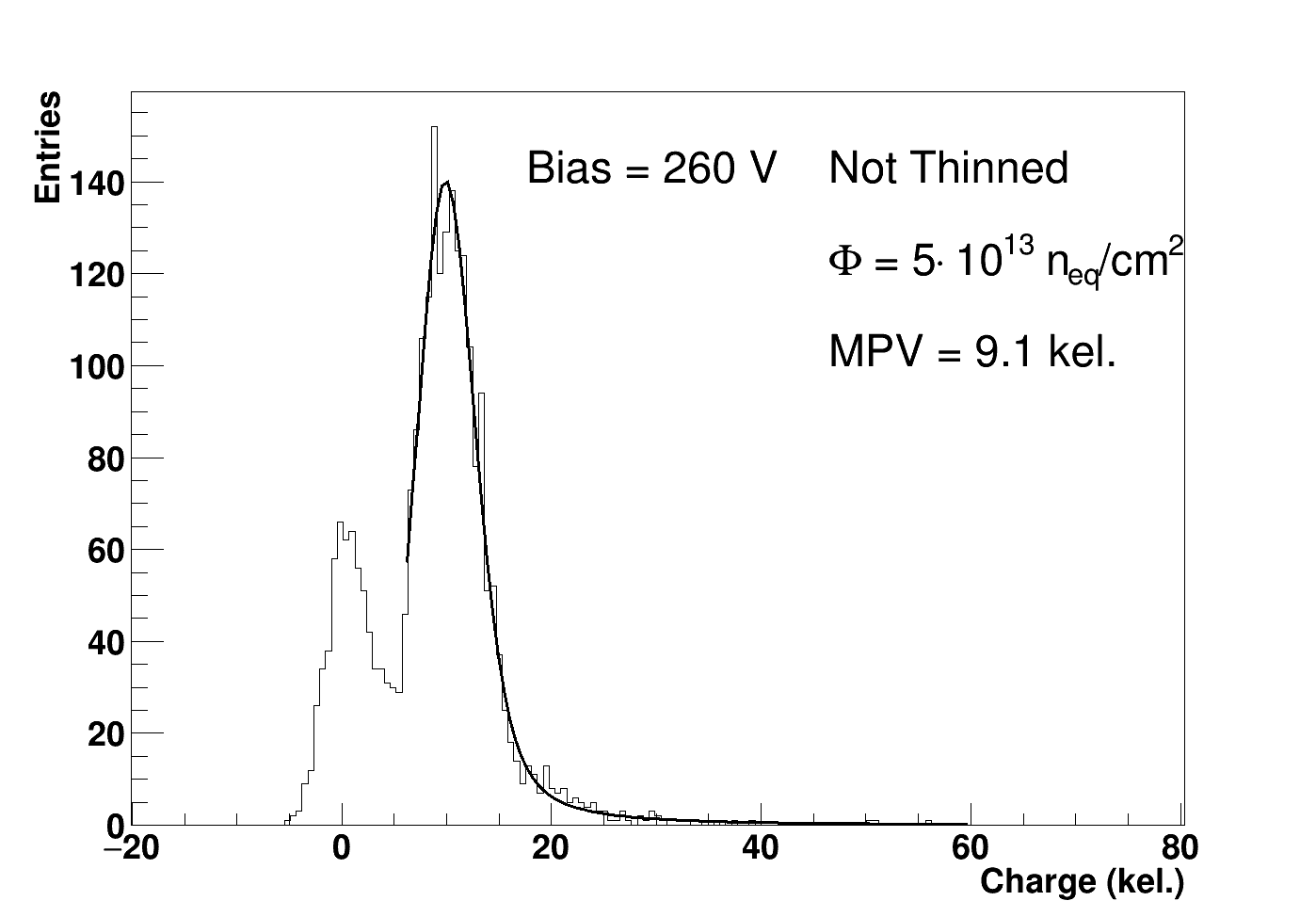} & \includegraphics[width=0.5\textwidth]{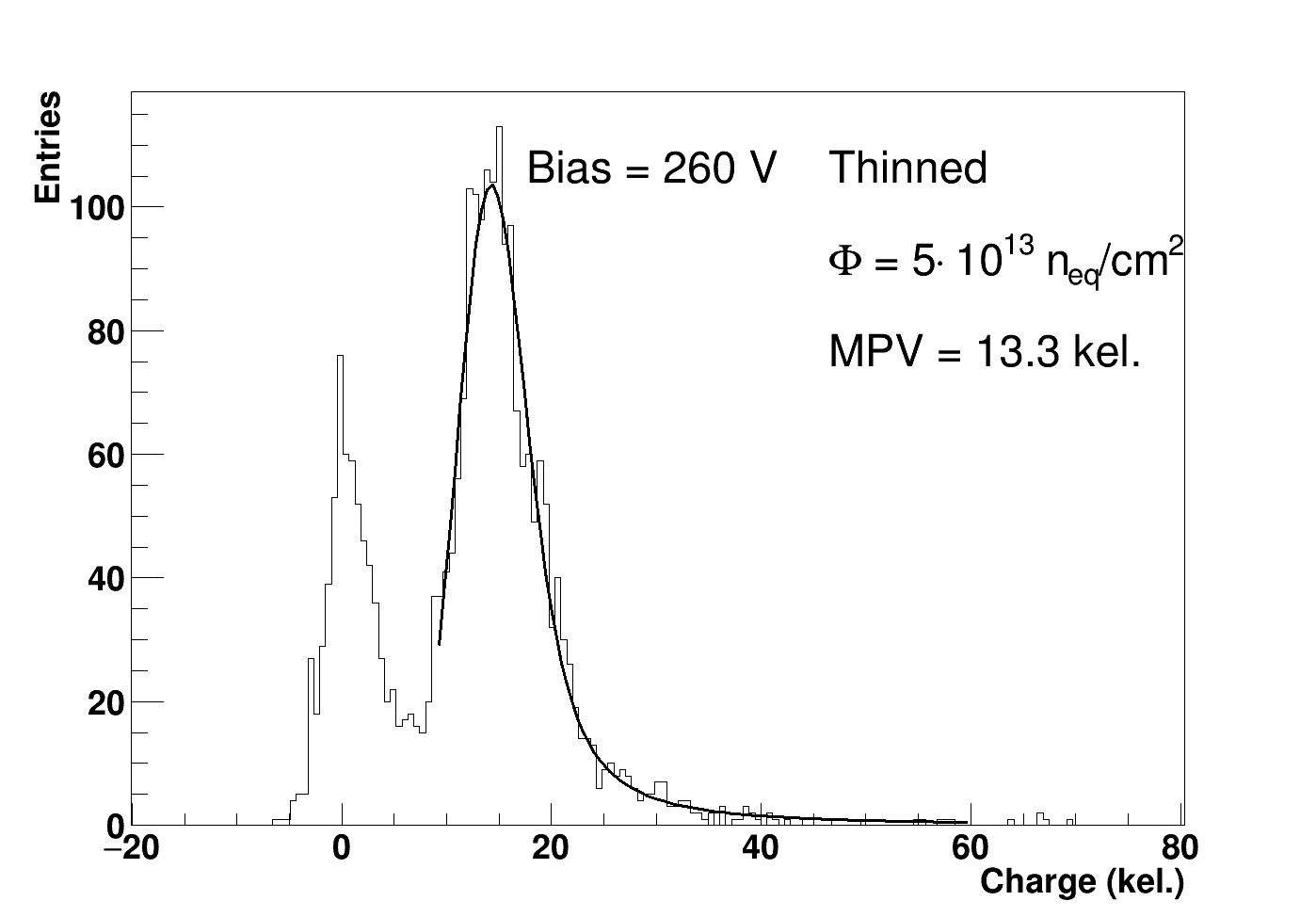} \\

  c)& d)

  \end{tabular}
  \caption{Spectrum of signals caused by electrons from $^{90}$Sr source for: unthinned (a) and thinned (b) device before irradiation, c) unthinned and d) thinned device after irradiation to 5$\cdot$10$^{13}$ n$_{\mathrm{eq}}$/cm$^2$.The charge is in kilo-electrons (kel.).}
\label{Sr90Spectra}
\end{figure}

For charge collection measurements with $^{90}$Sr source all pixels of device B (see Fig. \ref{LFstruct_photo}) were connected
together forming an effective pad detector.
Spectra measured before irradiation at 250 V for unthinned and thinned samples are shown in Fig. \ref{Sr90Spectra}a) and  \ref{Sr90Spectra}b).
A convolution of Landau and Gaussian functions is fitted to the spectra. The most probable value of the Landau function (MPV)
estimated from the fitted function is the measure of the collected charge. It can be seen in Fig. \ref{Sr90Spectra}a) that a larger MPV
was measured with unthinned devices. It corresponds to the charge of 20000 electrons, while with the thin sample a MPV of 13700 electrons was measured. This is a bit smaller than expected from Edge-TCT measurements of depletion depth shown in
Fig. \ref{ETCTunirrad}. A MIP releases $\sim$ 77 electrons per $\upmu$m of silicon with the largest probability but due to lower energy and scattering of $^{90}$Sr electrons about 10\% larger charge is expected.

\begin{figure}[!hbt]
\centering
\includegraphics[width=0.7\textwidth]{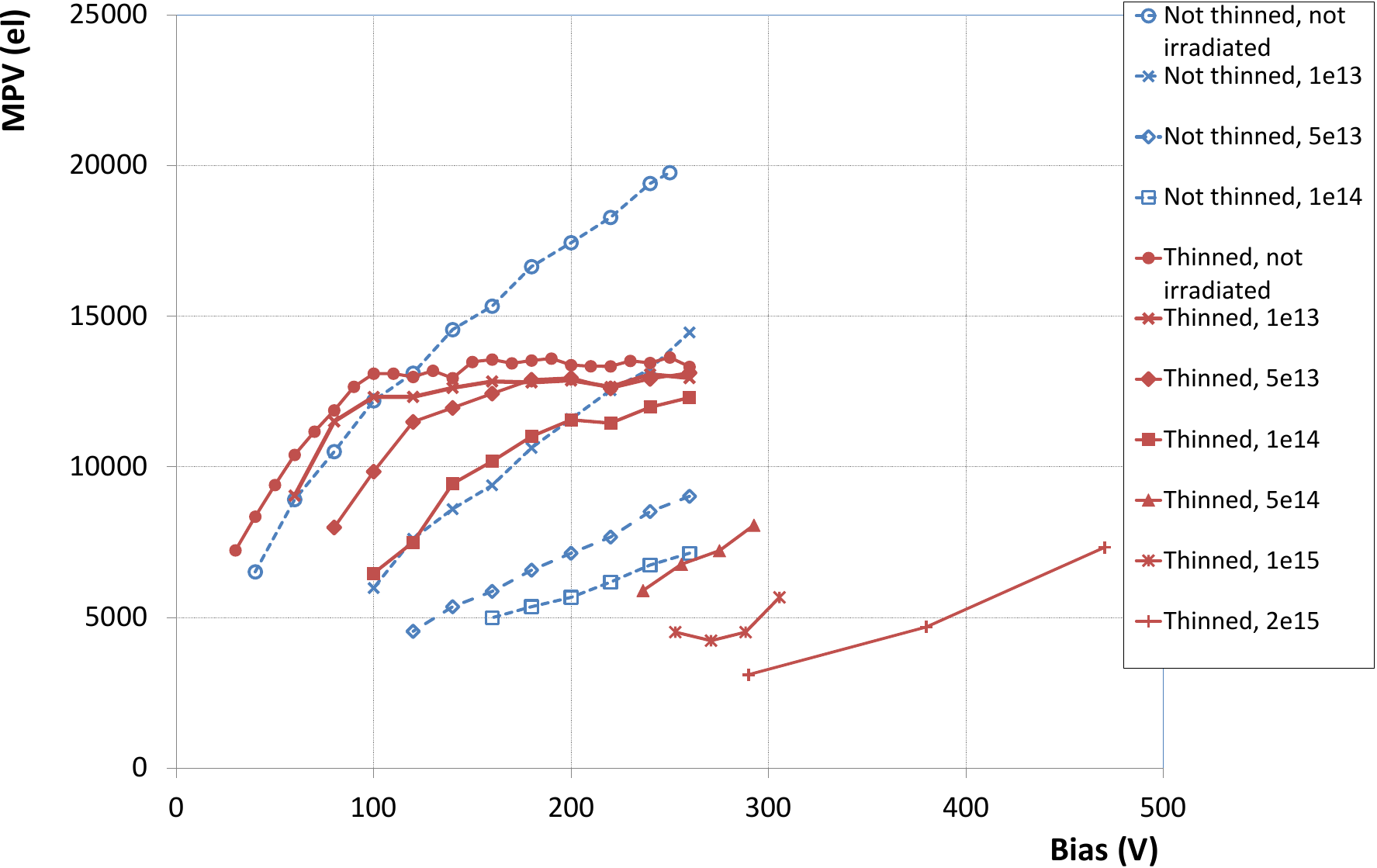} \\
a)\\ 
\includegraphics[width=0.7\textwidth]{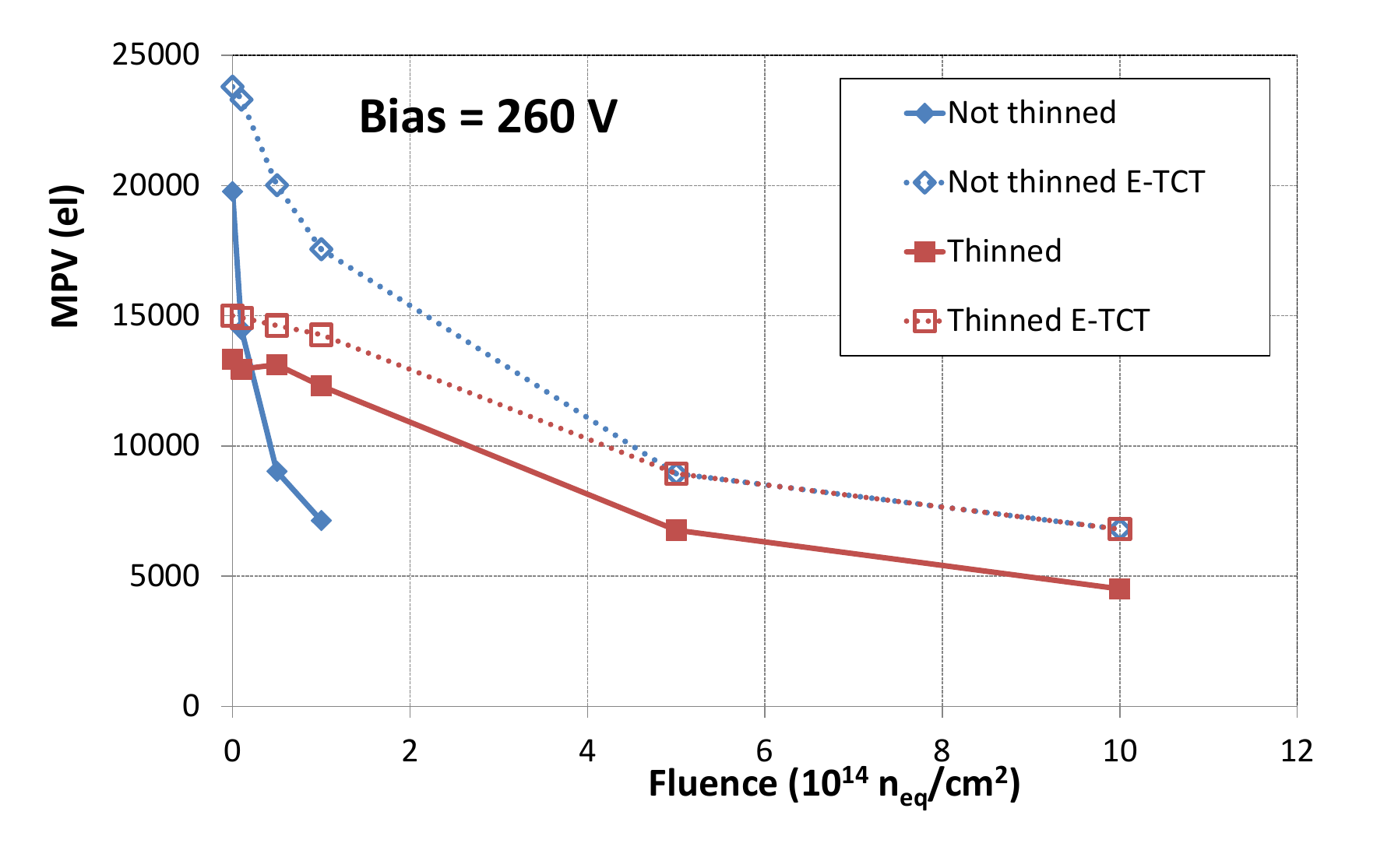} \\
b)
\caption{Collected charge vs. bias voltage for unthinned and thinned devices at different fluences is shown in figure a). Figure b) shows charge vs. fluence at 260 V bias. Shown are measured values and values calculated from depletion depth obtained from E-TCT.}
\label{MPVvsBias}
\end{figure}

Figure  \ref{Sr90Spectra}c) and  \ref{Sr90Spectra}d) show the spectra after irradiation with 5$\cdot$10$^{13}$ n$_{\mathrm{eq}}$/cm$^2$ and it can be now clearly observed that a significantly larger charge is measured with the thinned detector although the depletion depth obtained from Edge-TCT is significantly larger in the unthinned sample as shown in Fig. \ref{ETCTirrad}.

Depletion depth at  5$\cdot$10$^{13}$ n$_{\mathrm{eq}}$/cm$^2$ and 260 V bias voltage is 260 $\upmu$m in the unthinned sample and 180 $\upmu$m in the thinned sample which is fully depleted (see Fig. \ref{ETCTirrad}) at this bias voltage. Therefore, a fast electrons selected from  
$^{90}$Sr spectrum releases $\sim$ 22000 electrons MPV in the depletion region of the unthinned device and $\sim$ 15000 electrons in the
thinned one.
In Fig. \ref{Sr90Spectra}c) and d) it can be seen that the MPV of collected charge is 9000 electrons in the unthinned device and 13000 electrons in the thinned one. So while in the thinned device almost all charge is collected in the unthinned less than 50\% is seen.

Measurements were made also at other fluences and bias voltages as shown in Figure \ref{MPVvsBias}a). At the highest fluence bias voltage of over 350 V had to be applied to collect more than 4000 electrons. The devices could be biased to such a high voltage because breakdown performance improves with irradiation.

In thinned devices with backplane contact the collected charge after irradiation is about 15\% smaller than expected but it follows roughly the change of depletion depth measured with Edge-TCT. This is not the case for the unthinned devices without backplane where a large drop of collected charge is observed already at lowest fluence and much lower collected charge is measured than it is released in the depletion region by electrons from $^{90}$Sr. This is shown in Fig. \ref{MPVvsBias}b) where collected charge is plotted as a function of fluence at 260 V bias. Shown are measured values and values calculated from depletion depth estimated with Edge-TCT as shown in Figure \ref{ETCTirrad}. Charge values were calculated by multiplying the depletion depth with 84 electrons per
$\upmu$m (i.e. 10\% larger charge for $^{90}$Sr electrons than 77 electrons/$\upmu$m released by a MIP) and a factor roughly taking into account charge trapping. This factor is smaller than 15\% at highest fluence. The trapping factor was estimated with a simplified simulation using measured trapping probabilities from \cite{cindrop}.  The simulation was made with KDetSim simulation tool - a ROOT based library specialized
for simulation of charge drift in static electric field in silicon detectors \cite{kdetsim}. More detail about the simulation method can be found in \cite{hitixfab}.

%%Trapping of charge carriers during drift in the depletion region is not sufficient to explain this discrepancy.

The large difference between measured and TCT values for unthinned devices follows from different contacting schemes of the unthinned and thinned devices. As explained in section \ref{samples} in unthinned devices the substrate is biased via the p-type implant ring on top of the device while in the thinned devices the backplane was processed and metallised enabling contact to the substrate.

In case of top bias the drift paths of holes from the depletion region below the positively biased implant to the substrate contact are passing through the low field regions of the detector which does not significantly affect the charge collection before irradiation because trapping is negligible and the ohmic conductivity of the undepleted substrate is high enough to bring the zero weighting potential close to the border of the depletion region. In this case charge carriers drifting in the electric field of the depletion region cross at the same time a large part of the weighting field - the  condition necessary for high charge collection according to Ramo's theorem. This explanation is supported by measurements before irradiation in Fig. \ref{Sr90Spectra}a) and b) where the measured collected charge agrees with the charge deposited in the depletion depth measured with Edge-TCT.

The situation becomes different after irradiation because of trapping and increased ohmic resistivity of the substrate \cite{chilingarov}. In case of unthinned devices the zero weighting field potential is at the top of the sensor and therefore distant from the region of high electric field and carriers are trapped in the detector volume with low electric field before they reach the contact.
For the thinned devices the zero weighting potential electrode is at the backplane. In the case of full detpletion this is at the border of the depletion region and in the case of partial depletion significantly closer than in the unthinned and top biased device.  This reasoning implies that both - thinning and backplane processing improve charge collection in partially depleted detectors after irradiation. However it has to be repeated that measurements were made with all pixels of the structure connected together forming an affective pad detector in which weighting field is significantly different than in (small) pixel geometry. It can be expected that the effect of thinning and back biasing on charge collection would be significantly smaller in the pixel geometry where movement of carriers near pixel electrode contributes most to the collected charge.

\begin{figure}[!hbt]

  \begin{tabular}{cc}

 \includegraphics[width=0.5\textwidth]{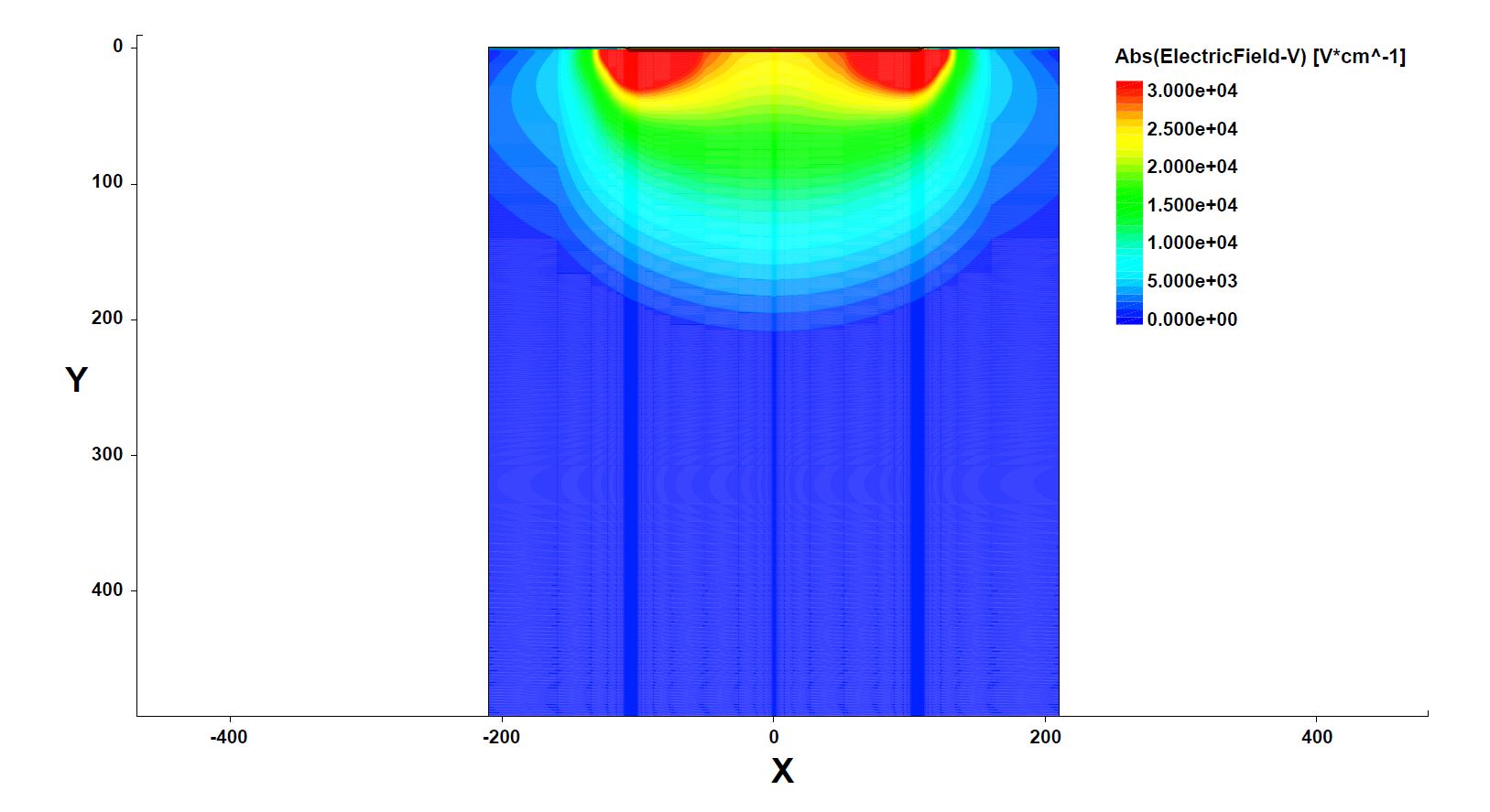} & \includegraphics[width=0.5\textwidth]{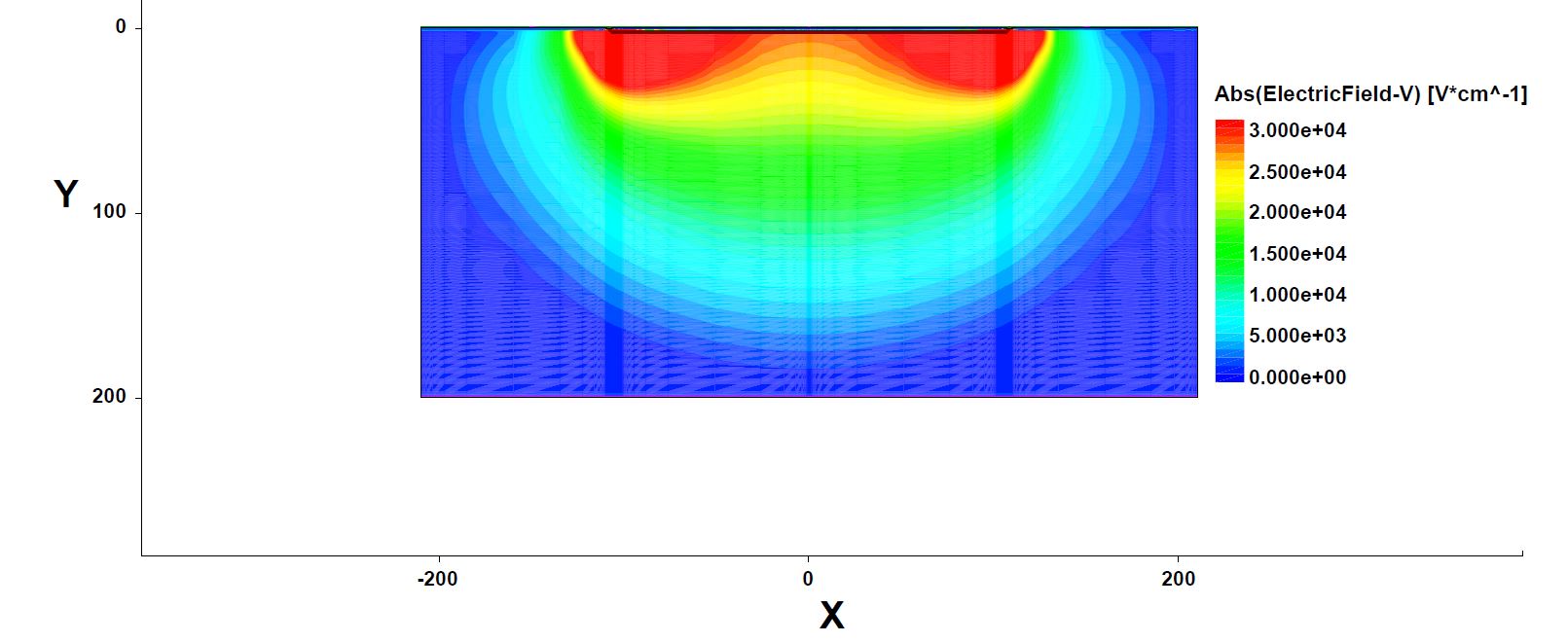}\\    
    a) & b) \\

 \includegraphics[width=0.4\textwidth]{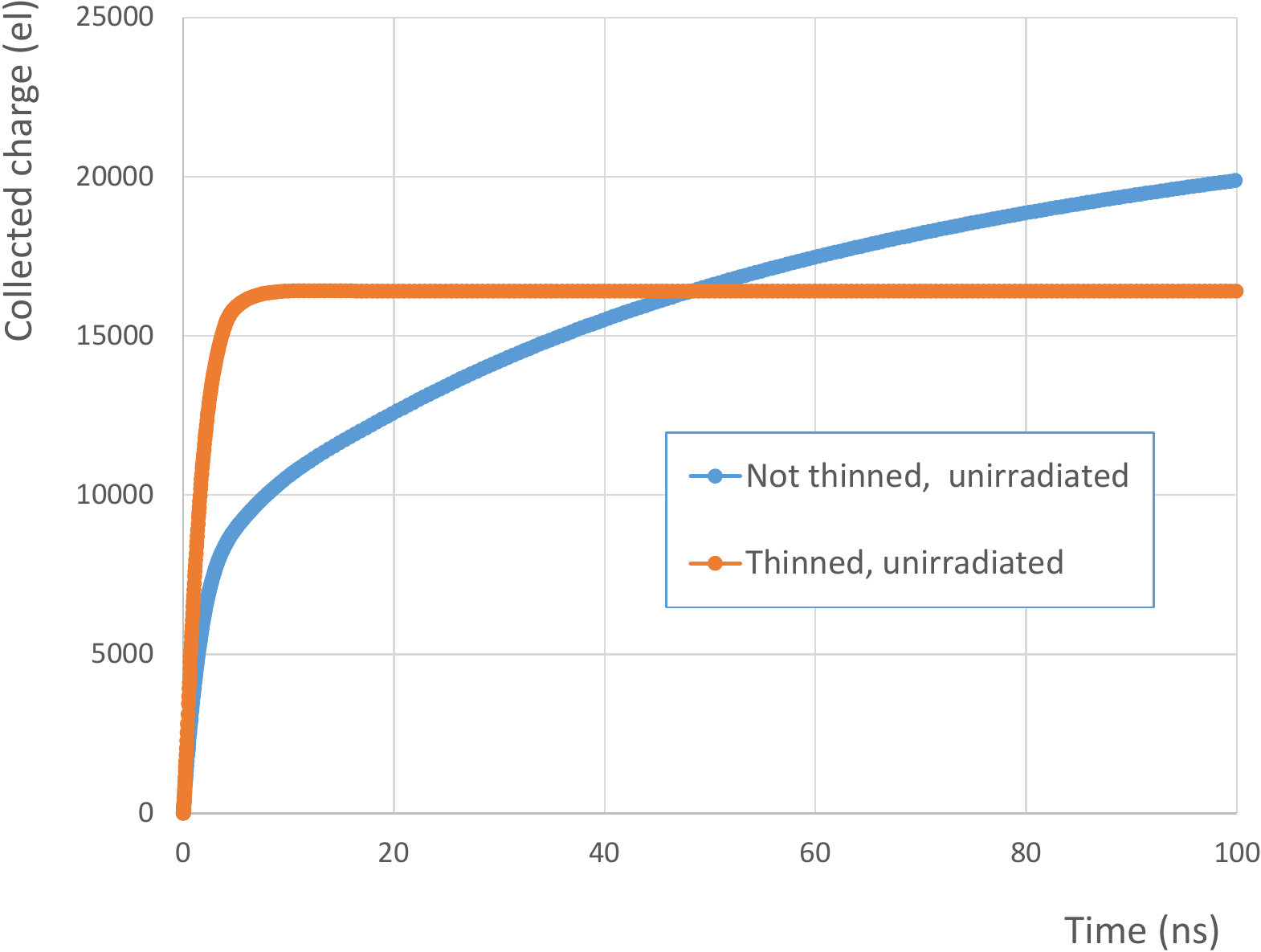} & \includegraphics[width=0.4\textwidth]{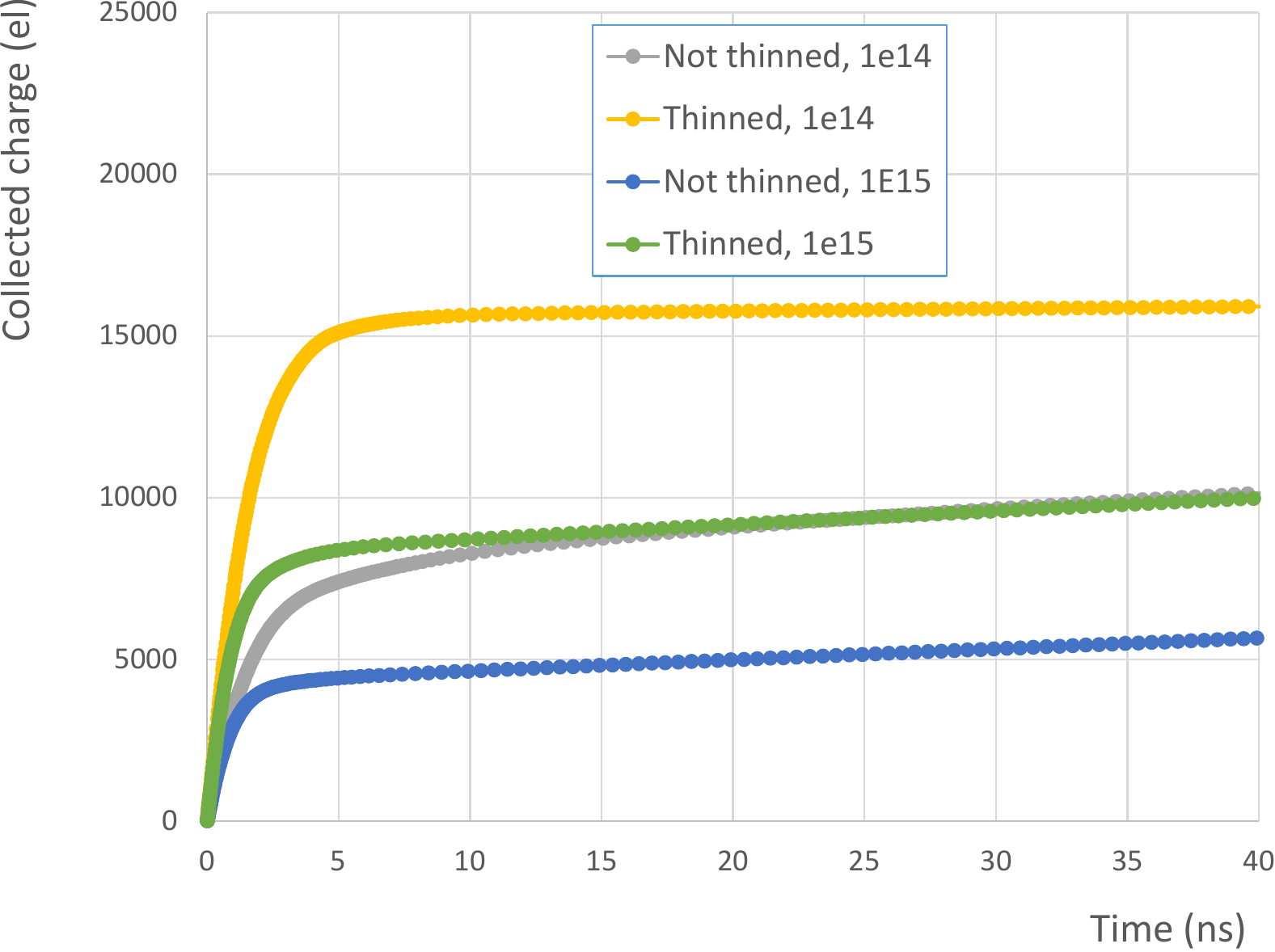}\\    
 c) & d) \\
 
  \end{tabular}
  \caption{Absolute value of electric field at 250 V bias voltage in a) unthinned structure biased from top and b) thinned structure with processed backplane before irradiation. Collected charge deposited along the centre of the structure (simulating a MIP) as a function of integration time before irradiation c) and after irradiation to two different fluences d), at bias of 250 V.}
\label{TCAD}
\end{figure}

\subsection{TCAD Simulations}

A TCAD simulation of a simplified structure was made with the aim to qualitatively support the above discussion about the differences of charge collection efficiencies for samples with and without processed backplane. A n-type implant was positioned in a p-type substrate with an initial resistivity of 3 k$\Omega$cm and two cases were considered: a 200 $\upmu$m thick device with substrate contacted through the backplane and a 700 $\upmu$m thick structure with substrate contacted on the area surrounding the implant on top.

Figures \ref{TCAD}a) and \ref{TCAD}b) show the distribution of the absolute value of electric field in the structure biased with 250 V before irradiation. Electric field in unthinned structure (Fig. \ref{TCAD}a) extends into substrate deeper than 200 $\upmu$m as expected for 3 k$\Omega$cm  resistivity while the thinned device is fully depleted.

A passage of a MIP was simulated by injecting 77 electron-hole pairs per $\upmu$m along the straight line through the centre of the structure. The current induced on the implant electrode by movement of the charge carriers in the electric field was calculated. Time integral of the induced current is the collected charge and it is shown in Figures \ref{TCAD}c) and \ref{TCAD}d) as a function of the integration time. It can be seen in Fig. \ref{TCAD}c) that before irradiation in the thinned sample with backplane bias all charge is collected rapidly while in the unthinned case a slow increase is seen after a steep rise. The slow component is the consequence of drift of carriers through the low field regions towards the contact on the top and also the consequence of the charge entering the field region by diffusion from the undepleted substrate. The simulation predicts that a larger charge would be collected with the unthinned device which is in agreement with measurements as can be seen in Figs.
\ref{Sr90Spectra}a) and \ref{Sr90Spectra}b).
On the other hand, in simulation equal charge in thinned and unthinned device is collected after $\sim$ 50 ns while 20000 electrons are collected in 100 ns in unthinned device. This is considerably slower than in measurements but it clearly illustrates the importance of backplane biasing for charge collection. The reason for the discrepancy with measurements was not studied in detail but it should be attributed to the simplifications used in the simulation: measurements were performed with an array of pixels while in simulation a single implant was considered. Also the slow charge collecting component in thick structure indicates that zero weighting field potential is not near the border of depletion zone which could be a consequence of underestimated conductivity of undepleted bulk.
The discrepancy between 13700 electrons, measured with thinned device before irradiation with the charge sensitive amplifier with 25 ns shaping time, and 16000 electrons collected in simulation in Figure \ref{TCAD}c) is partly a consequence of thickness of real device (see Fig. \ref{ETCTirrad}b) being only $\sim$ 180 $\upmu$m. 

The significance of thinning and back processing for charge collection efficiency can be much better seen after irradiation.
Properties of irradiated silicon were modeled in the simulation using the so called Perugia model \cite{perugia1, perugia2}. Collected charge as a function of integration time is shown in Fig. \ref{TCAD}d) for two fluences at 250 V bias. It can be seen that with the unthinned structure about a factor of 2 smaller charge is collected compared to the thinned sample. The value of measured collected charge is smaller than the charge in simulation however the factor of 2 difference is in agreement with measurements in Fig. \ref{MPVvsBias} at fluence of 1$\cdot$10$^{14}$ n$_{\mathrm{eq}}$/cm$^2$. At 1$\cdot$10$^{15}$ n$_{\mathrm{eq}}$/cm$^2$ collected charge in the unthinned device was too small to be measured with this measurement system. The cause for such a large difference between thinned and unthinned devices is the trapping of charge carriers in the low field regions of the unthinned sample, before they reach the contacts and so traverse the entire weighting field needed for full charge collection.

\section{Conclusions}

Measurements with a set of passive pixel detectors made in LFoundry 150 nm CMOS technology irradiated with neutrons were presented in this work. Two sets of devices were studied and compared: unthinned devices without processed backplane and biased through the implant on top and thinned devices with processed and metallised backplane enabling contact to the substrate. Depletion depth was measured with Edge-TCT and it was found that it increases with bias voltage following the square root dependence as expected in the case of abrupt junction and uniform doping concentration. From this dependence, the effective doping concentration $N_{\mathrm{eff}}$ was extracted.
It was found that in the range of fluences studied here $N_{\mathrm{eff}}$ increases linearly with fluence with somewhat larger introduction rate than usual for float zone silicon detector materials. This result is consistent with measurement in \cite{FirstLF}. The measurements confirm that also after irradiation with 2$\cdot$10$^{15}$ n$_{\mathrm{eq}}$/cm$^2$ depletion depths exceeding 50 $\upmu$m can easily be reached at bias voltages which can safely be applied in this technology.

Charge collection measurements with MIPs from $^{90}$Sr source revealed a large advantage of thinned devices with processed and metallised backplane after irradiation. Much larger collected charge was measured with irradiated thinned devices which brings a strong message that thinning (up to the level to remove the undepleted bulk) and backplane processing would improve charge collection of irradiated CMOS devices by modifying the electric and weighting field in the detector. However, it should be mentioned that measurements were made with effectively pad detectors so the effect in pixel geometry is expected to be smaller because of different weighting field.

Collected charge was measured up to the fluence of 2$\cdot$10$^{15}$ n$_{\mathrm{eq}}$/cm$^2$ and most probable value exceeding 5000 electrons was measured also at highest fulence if high enough bias voltage was applied. This amount of collected charge is sufficient for successful operation in the experiments at the HL-LHC.

\section{Acknowledgments}

The authors would like to thank the crew at the TRIGA reactor in Ljubljana for help with irradiation
of detectors.
Part of this work was performed in the framework of the CERN-RD50 collaboration. The authors acknowledge the financial
support from the Slovenian Research Agency (research core funding No. P1-0135 and project ID PR-06802). 
This project has received funding from the European Union's Horizon 2020 Research and Innovation programme under Grant Agreement no. 654168.

\end{document}